\documentclass[]{my_aastex701}

\usepackage{libertine}
\usepackage{libertinust1math}
\usepackage[T1]{fontenc}

\usepackage{datetime}
\newdateformat{mydate}{\THEYEAR\ \monthname[\THEMONTH] \THEDAY}

\shorttitle{High-spectral Resolution Center-to-limb Observations}
\shortauthors{Verma et al.}
\received{2025-10-29}
\accepted{2025-11-02}
\submitjournal{Research Notes of the American Astronomical Society}

\begin{document}


\title{\large High-spectral Resolution, Multi-wavelength Center-to-limb Observations of the Sun}

\author[orcid=0000-0003-1054-766X, sname='Verma']{Meetu Verma}
\affiliation{Leibniz Institute for Astrophysics Potsdam (AIP), 
    An der Sternwarte 16, 14482 Potsdam, Germany}
\email[show]{\textcolor{xlinkcolor}{mverma@aip.de}}  

\author[orcid=0000-0002-7729-6415, sname='Denker']{Carsten Denker}
\affiliation{Leibniz Institute for Astrophysics Potsdam (AIP), 
    An der Sternwarte 16, 14482 Potsdam, Germany}
\email{cdenker@aip.de}  

\author[orcid=0000-0002-0484-7634, sname='Pietrow']{Alexander G.\ M.\ Pietrow}
\affiliation{Leibniz Institute for Astrophysics Potsdam (AIP), 
    An der Sternwarte 16, 14482 Potsdam, Germany}
\email{apietrow@aip.de}  

\author[orcid=0000-0003-2059-585X, sname='Kamlah']{Robert Kamlah}
\affiliation{Leibniz Institute for Astrophysics Potsdam (AIP), 
    An der Sternwarte 16, 14482 Potsdam, Germany}
\email{rkamlah@aip.de}  

\author[orcid=0000-0002-8963-3810, sname='Petit dit de la Roche']{Dominique J.\ M.\ Petit dit de la Roche}
\affiliation{Leibniz Institute for Astrophysics Potsdam (AIP), 
    An der Sternwarte 16, 14482 Potsdam, Germany}
\email{dominiquepddlr@gmail.com}


\begin{abstract}
The center-to-limb variations (CLVs) of photospheric and chromospheric spectral lines were obtained in 2025 July and August using drift scans from the echelle spectrograph of the 0.7-meter Vacuum Tower Telescope (VTT) at the Observatorio del Teide (ODT) in Tenerife, Spain. This instrument can observe four spectral regions simultaneously, enabling multi-line spectroscopy with high spectral resolution of various activity features and the quiet Sun in the lower solar atmosphere. The initial results of H$\alpha$ observations demonstrate the diagnostic potential of drift scans obtained with a ground-based, high-resolution telescope. Data products include spectroheliograms and maps of physical parameters such as line-of-sight velocity, line width, and line-core intensity. The combination of the CLV from photospheric and chromospheric lines, as well as the wide range of formation heights of the selected lines, renders this dataset ideal for characterizing stellar and exoplanet atmospheres.
\end{abstract}


\renewcommand{\baselinestretch}{1.15}\selectfont


\section{Introduction} 

Solar spectroscopy of H$\alpha$ and other chromospheric lines has a long history. However, only a few comprehensive databases and archives contain such spectral data suitable for CLV studies. For instance, \citet{David1961} presented H$\alpha$ spectra of the quiet Sun obtained at seven different locations on the solar disk using the G\"ottingen Solar Tower and its Littrow spectrograph. One technique for full-disk spectroscopy is drift scanning, which involves letting the solar disk to move across the slit. \citet{Wittmann1976} used this method to study the CLV of various chromospheric lines in the quiet Sun. \citet{Lee2000} also presented drift scans obtained at the Big Bear Solar Observatory (BBSO), focusing mainly on small-scale dynamic H$\alpha$ features in the quiet Sun. Since October 2021, the Chinese H$\alpha$ Solar Explorer \citep[CHASE,][]{Li2022} has provided access to daily full-disk drift scans. \citet{Ellwarth2023} presented a CLV solar atlas containing 14 heliocentric positions. Recently, \citet{Hanassi-Savari2025} provided solar flux atlases for the quiet Sun using Sun-as-a-star observations from HARPS-N. Challan, a solar full-disk imaging spectroscopic telescope \citep{Yang2025}, scheduled to be installed at BBSO in 2025, will provide simultaneous H$\alpha$ and infrared Ca\,\textsc{ii} drift scans. The VTT, with its echelle spectrograph, has a long record of providing high-resolution spectra in the visible and near-infrared. Recent H$\alpha$ drift scans are presented to demonstrate the performance of the setup, which capitalizes on high-cadence and large-format scientific CMOS detectors.


\section{Observations} 

\begin{figure}
\centering
\includegraphics[width=\textwidth]{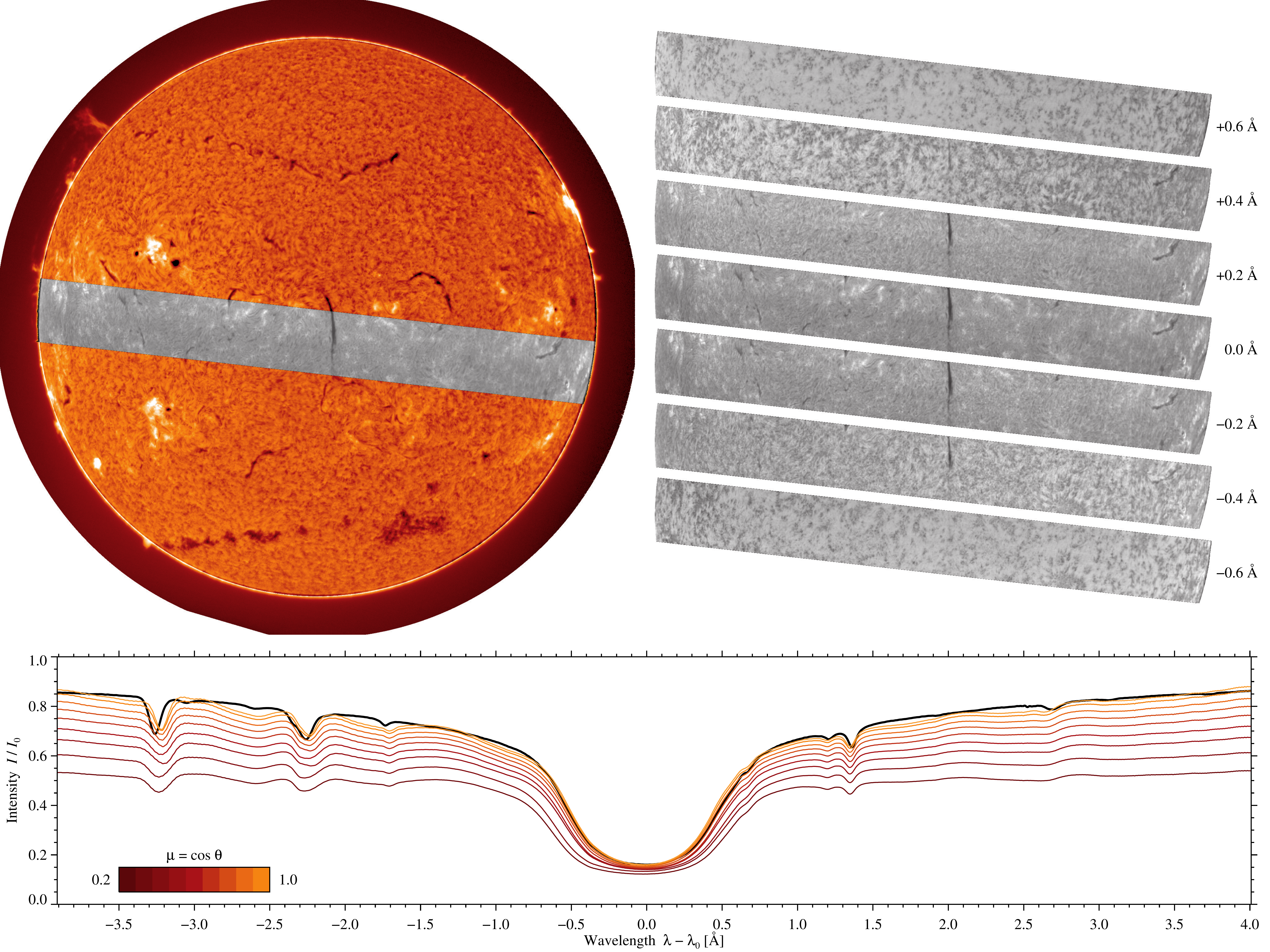}
\caption{GONG H$\alpha$ full-disk image obtained at 10:49:11~UT. The grayscale insert 
    is an H$\alpha$ line-core VTT drift scan image taken at 10:48:53~UT (\textit{top-left}). Spectroheliograms were taken at seven positions along the H$\alpha$ line, and the labels to the right indicate the wavelength offset from the line center (\textit{top-right}). The H$\alpha$ quiet-Sun spectra are shown as a function of the cosine of the heliocentric angle $\mu = \cos\theta$ in the interval $[0.2,\, 1.0]$ in steps of $d\mu = 0.1$. The thick black curve represents an FTS atlas reference profile (\textit{bottom}).}
\label{DRIFT_SCAN}
\end{figure}

From 2025 July~19 to August~1, fourteen drift scans were recorded with the Fast Multi-Line Universal Spectrograph (FaMuLUS) camera system, which included twelve spectral windows ranging from the near-ultraviolet to the near-infrared: Ca\,\textsc{ii}\,H $\lambda$396.9~nm, the red and blue parts of the Fraunhofer G-band $\lambda$430.7 nm, H$\beta^\ast$ $\lambda$486.1~nm, Cr\,\textsc{i}$^\ast$ $\lambda$578.2~nm, Na\,D$_1$ $\lambda$589.6~nm, H$\alpha^\ast$ $\lambda$656.3~nm, Fe\,\textsc{i}$^\ast$ $\lambda$709.0~nm, K\,\textsc{i} $\lambda$769.9~nm, Ca\,\textsc{ii} $\lambda$849.8~nm, Ca\,\textsc{ii} $\lambda$854.2~nm, and Ca\,\textsc{ii} $\lambda$866.2~nm. The four spectral regions that were simultaneously observed on 2025 July~22 are marked by an asterisk but only the H$\alpha$ data are exemplarily shown in Figure~\ref{DRIFT_SCAN}. The spatial scale along the slit is 0.36\arcsec\ pixel$^{-1}$, and the slit length is 216\arcsec\ corresponding to 600 pixels after 8-pixel binning. The spectra were recorded at a rate of 10~Hz. Full-disk H$\alpha$ images obtained at the ODT station of the Global Oscillation Network Group \citep[GONG,][]{Hill2018} were used to align the drift-scan data, accounting for translation, rotation, and pixel scale (1.398\arcsec\ pixel$^{-1}$) and skew ($\pm$5~pixel) in the scan direction. Approximately 800\,000 individual spectra were collected per scan and per spectral line. The nominal spectral resolving power of the spectrograph is ${\cal R} =$ 590\,900 at H$\alpha$. Each spectrum contains 1856 wavelength points after 4-pixel binning, and the wavelength range is 7.93~\AA\ with a reciprocal dispersion of 4.27~m\AA\ pixel$^{-1}$.

\section{Results and Conclusions} 

Figure~\ref{DRIFT_SCAN} shows the H$\alpha$ drift-scan data superposed on a co-temporal GONG H$\alpha$ full-disk image, illustrating image quality and spatial resolution. In addition, spectroheliograms at seven positions within the H$\alpha$ line demonstrate how the chromospheric features change their appearance with atmospheric height. Finally, the CLV of the H$\alpha$ line is plotted by averaging quiet-Sun spectra for different intervals of $\mu = \cos\theta$. The Kitt-Peak absolute disk-center intensity Fourier Transform Spectrometer \citep[FTS,][]{Neckel1999} atlas profile is provided for reference. As expected, the H$\alpha$ line becomes shallower and broader toward the limb \citep[e.g.,][]{Pietrow2023}. 

The unique property of VTT/FaMuLUS drift-scan data is that it consists of four spectral regions observed strictly simultaneously, which facilitates access to many other lines within the spectral range of the primary spectral line. Thus, the combination of the CLV, different formation heights, and bisectors allows for a fine-grained analysis of photospheric and chromospheric layers. In upcoming work, we will perform principal component analysis to denoise the spectra and conduct spectral inversions \citep{Dineva2020, Kuckein2021}. Ultimately, exoplanet studies \citep{Reiners2023} will benefit from this comprehensive analysis, which provides critical input for stellar activity simulations. Lastly, this research note highlights the VTT's ability for high-resolution, multi-wavelength spectroscopy. This builds on the recent study by \citet{Kamlah2025}, which demonstrated the VTT's potential for wide-field image restoration, bridging the gap between high-resolution and synoptic telescopes.


\begin{acknowledgments}
A German consortium led by the Institute for Solar Physics (KIS) in Freiburg operates the VTT, with the AIP and the Max Planck Institute for Solar System Research (MPS) in G\"ottingen as partners. This research has made use of the bibliographic services of NASA's Astrophysics Data System (ADS). AP was supported by grant PI~2102/1-1 from the Deutsche Forschungsgemeinschaft (DFG). MV acknowledges the support by a WISER grant from the Indo-German Science \& Technology Center (IGSTC-05373).
\end{acknowledgments}

\begin{contribution}
\textit{Contributor Roles Taxonomy} (\href{https://credit.niso.org/}{credit.niso.org}):
MV: conceptualization, software, validation, writing -- original draft;
CD: data curation, formal analysis, project administration, software, visualization;
ALL: investigation, writing -- review \& editing 
\end{contribution}

\setlength{\parindent}{0em}
\facilities{VTT \citep{vonderLuehe1998} -- GONG \citep{Hill2018}}
\software{sTools \citep{Kuckein2017}}

\renewcommand{\baselinestretch}{1.0}\selectfont


\newcommand{\BIBdecl}{\setlength{\itemsep}{1em}}
\bibliographystyle{spr-mp-sola}
\bibliography{apj-jour, drift_scan}

\end{document}